%% file: main.tex
\documentclass{l4dc2025}

\usepackage{algorithm}
\usepackage{algorithmic}
\usepackage{amsmath}
\usepackage{comment}
\usepackage{multicol}
\usepackage{bbm}
\usepackage{float}
\usepackage{booktabs}

\usepackage{acronym}

\usepackage{tikz,pgfplots}
\usepackage{textcomp}
\usepackage{makecell}
\usepackage{fontawesome}

\usepackage{graphicx}
\usepackage{epstopdf}
\usepackage{calc}

\usetikzlibrary{positioning,chains}
\usetikzlibrary{shapes.geometric}
\usetikzlibrary{calc}
\usetikzlibrary{trees}
\usetikzlibrary{decorations.pathreplacing,calc,tikzmark, automata, arrows}
\usetikzlibrary{backgrounds, fit}
\usetikzlibrary{positioning,arrows,petri,shapes.misc}
\usetikzlibrary{arrows.meta, bending, calc, fit, shapes, patterns}

\definecolor{TUMBlue}{RGB}{0, 101, 189}
\definecolor{TUMOrange}{RGB}{227, 114, 34}
\definecolor{TUMGreen}{RGB}{162, 173,   0}
\definecolor{TUMGray}{RGB}{51,  51,  51}
\definecolor{TUMRed}{RGB}{242,  12,  12}
\definecolorset{RGB}{PLOT}{}{%
	Brown, 153, 79, 0;%
	Black,   0,   0,   0;%
	Blue, 86, 180, 233;%
	DarkBlue, 0, 101, 189;%
	Orange, 230, 159, 0;%
	DarkOrange, 213, 94, 0;%
	Purple, 93, 58, 155;%
	Pink, 220, 38, 127;%
	Red, 227, 27, 35;%
	Yellow, 255, 195, 37;%
	BluishGreen, 0, 158, 115;%
	ReddishPurple, 204, 121, 167;%
	Olive, 162, 173, 0;%
	Gray, 153, 153, 153;%
	Cyan, 64, 176, 166;%
	DarkPurple, 136, 34, 85
}
\definecolor{TUMblue}{rgb}{0.00, 0.40, 0.74}
\definecolor{TUMgray}{rgb}{0.85, 0.85, 0.86}

\acrodef{rl}[RL]{reinforcement learning}
\acrodef{ppo}[PPO]{proximal policy optimization}
\acrodef{stl}[STL]{Signal Temporal Logic}
\acrodef{ode}[ODE]{ordinary differential equation}
\acrodef{pde}[PDE]{partial differential equation}
\acrodef{odes}[ODE]{ordinary differential equations}
\acrodef{pso}[PSO]{Particle Swarm Optimization}
\acrodef{gnn}[GNN]{Graph Neural Network}
\acrodef{nn}[NN]{Neural Network}

\DeclareMathOperator*{\argmax}{arg\,max}
\DeclareMathOperator*{\argmin}{arg\,min}
\def\always{\square}
\def\event{\lozenge}

\def\decrate{\gamma}  
\def\actrate{\nu}       
\def\discnt{K}        

\begin{document}

\title[Learning Biomolecular Models using Signal Temporal Logic]{Learning Biomolecular Models using Signal Temporal Logic}
\author{%
 \Name{Hanna Krasowski} \Email{krasowski@berkeley.edu}\\
 \addr University of California, Berkeley
 \AND
 \Name{Eric {Palanques-Tost}} \Email{ericpt@bu.edu}\\
 \addr Boston University, Boston%
 \AND
 \Name{Calin Belta} \Email{calin@umd.edu}\\
 \addr University of Maryland, College Park%
 \AND
 \Name{Murat Arcak} \Email{arcak@berkeley.edu}\\
 \addr University of California, Berkeley%
}

\maketitle

\begin{abstract}

Modeling dynamical biological systems is key for understanding, predicting, and controlling complex biological behaviors. Traditional methods for identifying governing equations, such as \acp{ode}, typically require extensive quantitative data, which is often scarce in biological systems due to experimental limitations. To address this challenge, we introduce an approach that determines biomolecular models from qualitative system behaviors expressed as \ac{stl} statements, which are naturally suited to translate expert knowledge into computationally tractable specifications. Our method represents the biological network as a graph, where edges represent interactions between species, and uses a genetic algorithm to identify the graph. To infer the parameters of the \acp{ode} modeling the interactions, we propose a gradient-based algorithm. On a numerical example, we evaluate two loss functions using \ac{stl} robustness and analyze different initialization techniques to improve the convergence of the approach.

\end{abstract}


\section{Introduction}
\acresetall
Biological models are central to understanding and manipulating 
processes such as immune system response, wound healing, and cancer. 
For example, a model of an organism with a chronic disease can provide detailed insights into its malfunctioning \citep{mose2019-ramodel}, or a model can be used as instruction to build a synthetic biological system that could be used to regulate an immune response \citep{nielsen2016-cello}. The two tasks, i.e., understanding and manipulating, are usually addressed by inference and synthesis, respectively. 
Biological network inference aims to uncover the underlying dynamics of an existing system by constructing models that align with usually quantitative data to replicate observed behaviors \citep{Delgado2019, peter2019-metabolomicinference}. 
Synthesis aims to build systems that comply with a desired behavior \citep{nielsen2016-cello}. 
Despite 
differing goals, both approaches share a common challenge: modeling dynamical systems that satisfy constraints derived from experimental data, qualitative observations, or desired outcomes.

Both inference and synthesis are challenging for the typically data-scarce problems in biology due to expensive real-world measurements. Constraining the model candidates by leveraging domain knowledge is therefore often a necessity, e.g., assuming the model is described by a set of differential equations \citep{Brunton2016-sindy} or adding prior knowledge about specific genes \citep{penfold2012-bayesianpriorinference}. 
Furthermore, the majority of available data of biological systems is qualitative and implicitly embedded in natural language; for example: “If the concentration of IL-6 and TGF$\beta$ exceeds a threshold for a certain period, the concentration of Th17 cells should increase, while the induction of Treg cells should decrease after a delay” \citep{martinez2018-tcellcytokinerole}. While such descriptions are rich in insight, they 
do not readily translate to computational procedures.
 
In this paper, we introduce a genetic algorithm that is combined with a gradient-based optimization to learn biomolecular models from \ac{stl} specifications. We propose to formalize qualitative information with \ac{stl} so that the information becomes computationally tractable. 
We represent the biomolecular system as a graph, where nodes are biological entities (i.e., cells, cytokines, or genes) and edges represent interactions (i.e., inhibition and activation) between them. The graph is transformed to an \ac{ode} model leveraging biological domain knowledge. Our genetic algorithm searches for the optimal graph structure while evaluating each population with respect to edge sparsity and satisfaction of \ac{stl} specifications. We optimize the unknown \ac{ode} parameters for each candidate of a population by employing a differentiable measure of \ac{stl} in the loss of a gradient-based algorithm.

Our contributions are: (1) We introduce an approach for learning interpretable and biologically-grounded \ac{ode} models from \ac{stl} statements, 
which is
applicable 
when only qualitative data is available. (2) The combination of a genetic algorithm with a gradient-based optimization allows for simultaneously identifying the network structure and unknown parameters for biomolecular models 
of intracellular or
intercellular processes. (3) The approach is suitable for both inference and synthesis, as the \ac{stl} specifications can describe existing or intended system behavior. (4) We propose and evaluate an adaptive loss function and a parameter initialization strategy to robustify convergence.

The remainder of this paper is structured as follows. In Section \ref{sec:related-work-and-contributions}, we contextualize our contributions with respect to existing literature. 
In Section \ref{sec:preliminaries} and Section~\ref{sec:problem-statement}, we define relevant concepts and formulate the main problem. 
We describe our algorithm to learn biomolecular models in Section \ref{sec:system-inference}, and show its effectiveness on a numerical example of a gene network in Section \ref{sec:experiments}. 
Discussions and conclusions are
in Section \ref{sec:discussion} and Section \ref{sec:conclusion}.

\section{Related Work} \label{sec:related-work-and-contributions}

\paragraph{Signal temporal logic for optimization.}
\ac{stl} \citep{maler2004-stllanguage} is commonly used in 
formal verification, where an \ac{stl} formula specifies a desired behavior, and its robustness measure is used to determine the degree 
to which
trajectories 
satisfy the formula. Using robustness as an objective function has been challenging due to the use of min and max functions, which are non-smooth. Some algorithms, such as those 
by \cite{bombara2021-decisiontree}, use gradient-free methods like decision trees, but they run into scalability issues for larger datasets. 
Recent work, e.g., \cite{ketenci2020-diffstl} and \cite{li2024-listlnn}, propose smooth approximations of robustness, which enable 
gradient-based optimization techniques using \ac{stl} robustness as objective. For this paper, we adopt one of such differentiable robustness formulations proposed by \cite{Leung2023}.

\paragraph{ODE inference for biological systems.} A common problem in systems biology is the inference of the parameters of the \acp{ode} modeling a system. Early methods, such as \cite{Varah1982-odeIdentOld}, 
estimate
the derivative of the observable times-series data
and use optimization techniques such as least squares to fit ODE parameters. 
However, these methods were not accurate for complex \acp{ode}. Recently, more nuanced methods have been proposed, such as Bayesian optimization for parameter fitting \citep{Huang2020-BayesianParamInference}. \cite{sun2024-odeparamestimation} 
have proposed combining \ac{pso} with reinforcement learning to estimate the parameters of stiff \acp{ode}, achieving state-of-the-art performance. 

In parallel, several methods were developed to infer not only the parameters but also the structure of \acp{ode} themselves. A key contribution was the publication of SINDy by \cite{Brunton2016-sindy}. This method consists of creating a large matrix with candidate terms for the \acp{ode} and then applying sparse regression to select the most relevant ones. However, a drawback of this method is the need to pre-select the possible terms of the \acp{ode}. \cite{raissi2018-sindyWithNN} addressed this issue using black-box neural networks as \acp{ode}, allowing 
flexibility to approximate the data at the expense of interpretability. 
Other  results with the same 
black-box approach
include \cite{chen2019-NeuralODEs}, which proposed modeling \acp{ode} with neural networks trained via backpropagation,
and 
\cite{raissi2017-pinns}, which 
introduced physics-informed neural networks
trained for solving \acp{pde}. 
More
recent results aim to reduce reliance on data and to 
maintain
interpretability of the final \ac{ode} model.
These include
work by \cite{Daryakenari2024-aiaristotle}, which provides support for grey-box systems where known equations are used when possible, and black-box neural networks are used otherwise. 

Compared to previous methods, our approach offers interpretable \acp{ode} that are well-suited for scenarios with limited data, where black-box methods may be untrainable or undesirable. 
Our method integrates qualitative expert knowledge effectively through \ac{stl} specifications. Furthermore, compared to SINDy, our representation of the network as a graph allows for better 
ability to model
complex interactions between species. Additionally, by using gradient descent, our method enables the optimization of any ODE parameters, not just the linear coefficients targeted by SINDy’s sparse regression.

\paragraph{Biological network synthesis.} 
Synthetic biology aims to design genetic circuits that satisfy a particular behavior and that can be engineered in a cell. One of the most common tools, Cello,  by \cite{nielsen2016-cello}, is a comprehensive framework for designing genetic circuits, translating specified 
behaviors  into DNA sequences for implementation in organisms like Escherichia coli. However, in Cello the desired behavior is specified with aggregation of logic functions and does not directly handle complex temporal dependencies that can be expressed with \ac{stl}. Several approaches for circuit synthesis have been proposed, such as those in \citep{golfeder2019-stlgrnsyn, gilles2010-stlgrnsyn2, Bartocci2013}, which use temporal logic languages, including \ac{stl}. These methods primarily constrain network design by ensuring compliance with expected temporal logic specifications. In contrast, our approach uses STL robustness as a differentiable loss function, allowing direct optimization of circuit parameters to achieve desired behaviors. This provides a more scalable and integrated framework for circuit synthesis.

\section{Preliminaries} \label{sec:preliminaries}

\subsection{Notation}

We use $\mathbf x: [0,T]\mapsto \mathbb{R}^N$ to denote the vector of species concentrations over the time interval $[0,T]$; $x\in \mathbb{R}^N$ 
is the concentration 
at a particular time and its $i$th entry $x_i$ is the concentration of species $i$. We  represent the interactions in a system with a directed graph $\mathcal{G} = (\mathcal{N}, \mathcal{E}, \mathcal{T})$, where $\mathcal{N}$ is the set of nodes, i.e., species, $ \mathcal{E}$ is the set of edges, i.e., interactions among species, and $\mathcal{T}$ is the set of edge types. An edge is specified with a tuple $e = (u, v, j) \in \mathcal{E}$ meaning that it goes from node $u$ to $v$ and has type $j$. The numbers of nodes and edges are defined by $|\mathcal{N}|$ and $|\mathcal{E}|$, respectively. We denote the set $\mathcal{F} = \mathcal{N} \times \mathcal{N} \times \mathcal{T}$ to represent all edges and edge types that could exist in a graph. We also define a mapping $\Psi: \mathcal{G} \mapsto \mathcal{D}$ that converts the information in $\mathcal{G}$ into a dynamical system represented by \acp{ode} with parameters $\theta \in \mathcal{P}$ and external inputs $a$. The mapping associates each edge $e$ with specific terms such as the ones presented in Section~\ref{sec:math-modeling}, depending on the type of edge. For instance, an edge $e$ of type $j=1$ could be modeled with Eq.~\eqref{eq:active-hill} between two species with concentrations $z$ and $y$.

\subsection{Signal Temporal Logic} 
We use temporal specifications defined by \ac{stl} formulas; see  \cite{maler2004-stllanguage} for the formal 
syntax and semantics. 
\ac{stl} formulas 
have
three components: (1) predicates 
$\mu:= g(x) \geq 0$, with $g:\mathbb{R}^n\rightarrow\mathbb{R}$; (2) Boolean operators, such as $\neg$ (\emph{negation}), $\land$ (\emph{conjunction}), and $\implies$ (\emph{implication}); and (3) temporal operators, such as $\event$ (\emph{eventually}) and $\always$ (\emph{always}). Given as \ac{stl} formula $\phi$, $\event_{[t_1,t_2]}\phi$ is true if $\phi$ is satisfied for at least one time point $t\in[t_1,t_2]$, while $\always_{[t_1,t_2]}\phi$ is true if $\phi$ is satisfied for all time points $t\in[t_1,t_2]$. 

The semantics of \ac{stl} is defined over functions of time, or signals, such as trajectories produced by a system of \acp{ode}. \ac{stl} has both qualitative (Boolean) and quantitative semantics. In the Boolean semantics,  a trajectory $\mathbf x$ either satisfies a formula $\varphi$  (written as ${\mathbf x} \models \varphi$) or violates it. 
The quantitative semantics \citep{donze2010robust} is defined using a function $\rho(\mathbf x,\varphi)$, called robustness, which measures how strongly formula $\varphi$ is satisfied by $\mathbf x$.
In particular,
$\rho(\mathbf x, \varphi) \geq 0$ if and only if ${\mathbf x} \models \varphi$. The time horizon of an \ac{stl} formula $\varphi$, denoted as $hrz(\varphi)$, is the minimum amount of time required to decide the satisfaction of $\varphi$. For simplicity, we assume that $T=hrz(\varphi)$. For example, the quantitative semantics for the conjunction of formulas $\varphi_1$ and $\varphi_2$ is defined as \citep{Leung2023}:

\begin{equation}\label{eqn:robustness-conj}
    \rho(\mathbf{x}, \varphi_1 \land \varphi_2) = \min(\rho(\mathbf{x}, \varphi_1), \rho(\mathbf{x}, \varphi_2)).
\end{equation}
However, this definition is non-smooth due to the $\min$ operation. The smoothly differentiable version of the conjunction introduced in \citep{Leung2023} is:

\begin{equation}\label{eqn:robustness-conj-diff}
    \rho(\mathbf{x}, \varphi_1 \land \varphi_2) \approx \widetilde{\min}(\underbrace{\rho(\mathbf{x}, \varphi_1)}_{ = s_1}, \underbrace{\rho(\mathbf{x}, \varphi_2))}_{ = s_2} = \frac{s_1\exp(-\nu s_1) + s_2\exp(-\nu s_2)}{\exp(-\nu s_1)+\exp(-\nu s_2)}.
\end{equation}

\subsection{Mathematical Modeling of Biological Systems} \label{sec:math-modeling}

Typical models in systems biology  employ \acp{ode} with 
equations that describe the dynamics of molecular and cellular interactions; \cite{murray01,keener04,keshet05,alon2019introduction}. Among them, Hill equations are widely used to describe the non-linear binding and regulation of biological molecules. 
As an example, Eq. \eqref{eq:active-hill} below indicates positive regulation of a molecule, whose concentration is denoted by $y$, by another molecule with concentration $z$,
while Eq. \eqref{eq:inhib-hill} describes  negative regulation. In both cases, $\actrate$ represents the maximum activation/inhibition rate, $\discnt$ represents the affinity between 
the two molecules, and 
the {\it Hill coefficient} $n$ determines the steepness of the curve, with larger parameters tending towards a step-like behavior. The term $-\gamma \cdot y$ in each equation models degradation at a rate proportional to the concentration: 
\vspace{-1cm}
\begin{multicols}{2}
  \begin{equation} \label{eq:active-hill}
    \dot{y} = -\gamma \cdot y \ + \ \actrate \cdot \frac{\discnt \cdot z^n}{1 + \discnt \cdot z^n}
\end{equation}
\break
  \begin{equation} \label{eq:inhib-hill}
    \dot{y} = -\gamma \cdot y \ +  \ \actrate \cdot \frac{1}{1 + \discnt \cdot z^n}.
  \end{equation}
\end{multicols}

Note that a differential equation model can also include a combination of positive and negative regulation terms. There are other modeling paradigms in systems biology, including those that account for stochasticity, such as Chemical Master Equation models and their diffusion approximations leading to the Chemical Langevin Equation; see, e.g., \cite{vanKampen2007}. As a starting point, in this paper we focus on ODE models with standard activation and inhibition terms listed above due to their computational tractability.

\section{Problem Statement and Approach} 
\label{sec:problem-statement}

Given species of a system that are represented as nodes $\mathcal{N}$ of a graph $\mathcal{G}$, an observable or intended behavior for that system in the form of an \ac{stl} formula $\varphi$, and a mapping $\Psi$ that converts graphs $\mathcal{G}$ to dynamical systems $\mathcal{D}$ with parameters $\theta$, we want to find the edges of the graph $\mathcal{E}^* \in \mathcal{F}$ and the parameters of its respective dynamical system $\theta^* \in \mathcal{P}$ such that the robustness of the traces $\mathbf{x}$ generated by $\mathcal{D}$ is maximized. Mathematically:

\begin{equation}
\begin{aligned}
    \theta^*, \mathcal{E}^* &= \argmax_{\theta \in \mathcal{P},\, \mathcal{E} \in \mathcal{F}} \rho(\mathbf{x}(\theta, \mathcal{E}), \varphi)  
\end{aligned}
\end{equation}

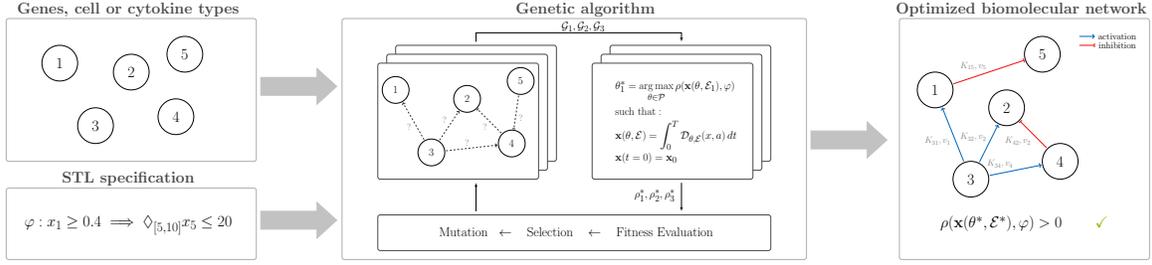
\begin{figure}[tb]
    \centering
    \input{figures/overview_fig.tex}
    \caption{Model learning approach: Given a set of species and a formalized specification, a genetic algorithm searches for the optimal topology while the parameters for each candidate graph are optimized based on the robustness of the formalized specification. The result is a specification-compliant biomolecular model with optimized parameters that is modeled by \acp{ode}.}
    \label{fig:headfigure}
\end{figure}

Our approach to the above problem is illustrated in Figure~\ref{fig:headfigure}. A major challenge is that the parameters $\theta$ can only be optimized and evaluated for a given edge set $\mathcal{E}$. Thus, we develop a genetic algorithm where the fitness function is calculated based on the topology of the graph and the robustness of the optimized system. Additionally, we propose two extensions: a pre-optimized parameter initialization to decrease the sensitivity to the initialization of the parameters $\theta$, and an adaptive loss for conjunction \ac{stl} specifications that switches to an additive robustness when the \ac{stl} specification is satisfied.

\section{Biomolecular Model Learning} \label{sec:system-inference}

In our approach, we consider two possible edge types: inhibition and activation. Our mapping $\Psi$ relates inhibition and activation edges to Eq.~\eqref{eq:inhib-hill}~and~\eqref{eq:active-hill}, respectively. The mapping also 
incorporates the degradation term of each species, which, intuitively, in the graph would be a self-loop around each node. 
In particular, we define the mapping between $\mathcal{G}$ and the dynamical system $\mathcal{D}$ with the \acp{ode} $\mathcal{D}_{\theta, \mathcal{E}} (x, a)$:

\begin{align}
    \dot{x}_u =& - \decrate_u x_u + \actrate_u \frac{\mathbbm{1}_d + \mathbbm{1}_r \sum_{\{e | \mathtt{sink}(u, e) \land \mathtt{type}(a, e) \}} \discnt_{e} \, x_v^n}{1+ \sum_{\{e | \mathtt{sink}(u, e)\}} \discnt_{e}\, x_v^n} + a_u,
\end{align}
Where $a \in \mathbb{R}^N$ is the external input and $a_u$ is the external input of node $u$, $\mathbbm{1} \in \{0,1\}$ denotes a binary variable,  $\mathbbm{1}_d =1$ if there are solely inhibition edges incoming to $u$, and $\mathbbm{1}_r =1$ if there is at least one activation edge incoming to edge $u$. The function $\mathtt{sink}(u, e)$ evaluates to true if $u$ is a sink node for the edge $e$, and the function $\mathtt{type}(j, e)$ that evaluates to true if the edge $e$ is of type $j$. Based on this formulation, our approach aims to find the set of edges $\mathcal{E} \in \mathcal{F}$ and parameters $\theta \in \mathcal{P}$, i.e., decay rates $\decrate$, maximum activation/inhibition rate $\actrate$, and affinity factor $\discnt$, that maximize the robustness $\rho$ of the \ac{stl} specification $\varphi$.

\subsection{Network Learning}
We learn the biological network structure with a genetic algorithm. We start from an initial population~$pop_0$ where we randomly create edges while randomly sampling edge types. We bias the sampling towards sparsity by sampling between $|\mathcal{N}| - 1$ and $\frac{|\mathcal{N}|^2}{4}$ edges. In case not all species
are connected after sampling, we add the minimum number of edges necessary to connect subgraphs to one graph and randomly sample an edge type. For each candidate graph, we optimize the parameters $\theta$ with respect to the \ac{stl} specifications as detailed in Section~\ref{subsec:param_opt}. The selection function identifies the top $k$ performers based on the fitness function that considers the optimized robustness $\rho^*$ and the number of edges:
\begin{equation}
    \mathcal{L}_{GA} (\rho^*, |\mathcal{E}|, \rho_{best}) = \frac{\rho^*}{\rho_{best}} + 1 - \frac{|\mathcal{E}| - |\mathcal{N}| + 1}{|\mathcal{N}|^2 - |\mathcal{N}|},
\end{equation}
where $\rho_{best}$ is the best-observed robustness up to the current generation.
For the next population, we mutate the top population with three mutation types with probability $\alpha$ each: removing one edge, adding one edge, resampling the edge types. Note that we post-process the edge removal mutation to ensure the graph is still fully connected. 

\subsection{Parameter Optimization}\label{subsec:param_opt}
The differentiable \ac{stl} specification enables us to use the robustness of a trace as the loss function in our optimization,
summarized in Algorithm~\ref{alg:optimization}: First, a trace for the system with the current parameters is computed. The loss $\mathcal{L}_\rho$ is the negative robustness of the trace. Then, the parameters are updated with a gradient descent method, such as Adam. Note that we approximate the continuous-time differential equation with a discrete-time Euler step update, 
as this allows us to efficiently backpropagate the gradient. We propose two extensions to our algorithm to improve efficiency and effectiveness: a robust parameter initialization strategy and an adaptive loss function.

\begin{algorithm}[tb]
\caption{Gradient-based parameter optimization with \ac{stl} specifications}
\label{alg:optimization}
\begin{algorithmic}[1]
    \STATE \textbf{Initialize:} \acp{ode} $\mathcal{D}_{\theta, \mathcal{E}}(x, a)$, parameters $\theta = \discnt_i \forall i \in \{1, ..., |\mathcal{E}|\}, \actrate_j, \decrate_j, j \in \{1 , ... , |\mathcal{N}|\}$, STL specification $\varphi$, initial state $\mathbf{x}_0$, parameter set $\mathcal{P}$, hyperparameters $\lambda, \epsilon$

    \WHILE{$\lnot$ converged}
         \STATE Compute trace $\mathbf{x}$ with $\mathcal{D}_{\theta, \mathcal{E}}(x, a)$ and $\mathbf{x}_0$
         \STATE Compute loss $\mathcal{L}_\rho = - \rho_n(\mathbf{x})$
         \STATE Update parameters $\theta \gets \theta - \lambda \nabla_{\theta} \mathcal{L}_\rho $
         \STATE Project parameters on parameter set $\theta \gets \mathtt{clip}(\theta,\mathcal{P}) $
         \STATE converged $\gets$ \textbf{true} if $| \rho_{n} - \rho_{n-1} | \leq \epsilon$
    \ENDWHILE
\end{algorithmic}
\end{algorithm}

\paragraph{Parameter initialization.}
Initialization of the \ac{ode} parameters plays a critical role in the convergence of the algorithm, especially when more complex \ac{stl} specifications are provided. We suggest performing a parameter search to initialize the parameters with the lowest loss function possible. That is, given an initial set $\Theta_0$ of possible initial parameters, we take the initial parameters that lead to the minimum loss function.

\begin{equation}
    \theta_0 = \argmin_{\theta \in \Theta_0} \mathcal{L_{\rho}(\mathbf x(\theta))}
\end{equation}
The set $\Theta_0$ can be created using a grid or from random samples. Here, we use Sobol sequences to fill the parameter space within some prefixed bounds. 

\paragraph{Adaptive loss function.}
Often, the \ac{stl} specification will be based on multiple qualitative statements that have to hold simultaneously. This translates to a conjunction of multiple subspecifications reflecting each of the statements. The robustness of a conjunction is the minimal robustness of the subspecifications (see Eq.~\eqref{eqn:robustness-conj}). Thus, the parameters are optimized mostly with respect to the least robust specification. While this objective is meaningful as long as the specification is not fulfilled with the current parameters, it is not clear if this is the best choice once all subspecifications are fulfilled, especially since it is difficult in practice to scale all predicates to one range. Additionally, the least robust subspecification might not be optimizable beyond a specific threshold, while for the other subspecifcations there might be still room for improvement without decreasing the robustness of the least robust subspecification. To address these concerns, we propose to use the following adaptive loss function:

\begin{equation}
    \mathcal{L}_{S} = \begin{cases} 
      \mathcal{L}_{\rho}& \rho \leq 0 \\
       - \sum_{j=1}^{J} \rho_j - \beta \sum_{j=1}^{J} \log(\rho_j), & \rho > 0 
   \end{cases}
\end{equation}
where $\mathcal{L}_\rho$ is the original loss as formulated in Algorithm~\ref{alg:optimization}, $\rho_j$ denotes the robustness of subspecification $\varphi_j$ and the \ac{stl} specification consists of $J$ subspecifications. 
The first sum aims to maximize the robustness gain while the second term penalizes stepping toward the satisfaction boundary, i.e., $\rho = 0$. 

\section{Numerical Example} \label{sec:experiments}
Our biomolecular model learning can be applied to multiple biological networks such as cell-cell and cell-cytokine networks, or gene regulatory networks. For our evaluation, we focus on 
gene networks, in which signal propagation is 
an important phenomenon \citep{Matsuda2012}. An informal specification of a propagation system could be: "Given protein X, produce protein Y within a short time interval". For our numerical evaluation, we investigate such a specification for a system with six species, i.e., $|\mathcal{N}| =6$, where if species $1$ or $2$ is above a threshold, two other species $3$ and $4$ increase in concentration, while never exceeding a maximum concentration. In particular, we consider the \ac{stl} specification:
\begin{align}
    \phi: &\left(x_1 \geq 0.2 \land x_2 \geq 0.3 \implies (\event_{[0,10]} \,  x_3 \geq 0.5) \land 
    (\event_{[0,10]} \always \,  x_4 \geq 0.9) \right) \, \land \notag \\
    &\left( x_4 \geq 0.6 \implies \event_{[0,20]} \always \, x_3 \leq 0.3\right) \notag\\
    &\land \left(\always x_1 \leq 1.5 \right)\land \left(\always x_2 \leq 1.5 \right)\land \left(\always x_3 \leq 1.5 \right)\land \left(\always x_4 \leq 1.5 \right).
\end{align}

The initial state is specified as $\mathbf{x}_0 = [0, 0, 0, 0, 1, 1]$, the propagation signal is $u=[0.5, 0.5,0, 0, 0, 0]$. We restrict all, i.e., $p$, parameters in the interval $\mathcal{P} = [0.001, 1]^p$ and set the Hill factor to $n=2$. The parameters for the evolutionary algorithm are top population $k=5$, mutation rate $\alpha=0.7$, the initial population is $pop_0 = 15$, and we search for $10$ generations. For the gradient-based optimization, we use Adam for gradient descent and set the learning rate to $\lambda= 0.04$, the convergence threshold $\epsilon=0.00001$, the weight in the adaptive loss to $\beta =1$. For the parameter initialization, we sample 20 times using Sobol sequences from the parameters space $\mathcal{P}$ to identify the best initial parameters.

\subsection{Results}
\begin{figure}[tb]
    \centering
    \input{figures/result_graph.tex}
    \includegraphics[width=0.47\textwidth]{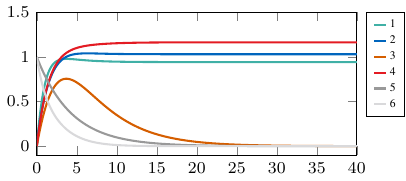}
    \caption{Gene network with highest robustness. Left: Graph including optimized affinity factors while the decay rates are $\decrate = [0.77, 0.81, 0.20, 0.63, 0.22, 0.93]$ and the activation/inhibition rates are $\actrate = [0.51, 0.93, 1, 0.73,0.22, 0.91]$. Right: Trajectory for gene network showing satisfaction of the specification $\phi$.}
    \label{fig:prop_sys_qualitative_results}
\end{figure}

We observe that in 9 of the 10 generations, our genetic algorithm finds at least one system with robustness above $0.15$. Figure~\ref{fig:prop_sys_qualitative_results} shows the gene network with the highest robustness, 
$0.251$, 
obtained by the genetic algorithm. The species 5, which starts from a high concentration, activates species 3, which inhibits species 4. This relates to the second subspecification. To fullfill the first subspecification, species 5 regulates species 3 and 4. Thus, the implication does not necessarily lead to a path between the species for the given initial state and propagation signal. 

\begin{table}[tb]
    \centering
        \begin{tabular}{l c c c c}
    \toprule
         &  \multicolumn{2}{c}{$\mathcal{L_{\rho}}$} &  \multicolumn{2}{c}{$\mathcal{L_{S}}$}\\ \cmidrule(l{2pt}r{2pt}){2-3} \cmidrule(l{2pt}r{2pt}){4-5}
         &  uniform & Sobol & uniform & Sobol \\ \midrule
         $\rho(\mathbf{x},\phi)$  & $0.167\pm 0.087$  &  $0.194\pm 0.093$ & $0.116\pm 0.067$& $0.162\pm 0.067$\\
         $\sum_i \rho_i(\mathbf{x},\phi_i)$   & $0.558\pm 0.217$ &$0.643\pm 0.238$& $0.628\pm 0.137$ & $0.719\pm 0.179$\\
         steps to convergence & $76.3\pm 5.5$ & $65.4\pm 19.2$&  $73.4\pm 14.5$ & $73.8\pm 12.2$\\
         steps to satisfaction & $16.6\pm 15.4$ & $8.6 \pm 12.6$&  $17.2\pm21.6$ & $9.1\pm15.6$\\  \bottomrule
    \end{tabular}
    \caption{Comparison of parameter optimization with different initialization strategies (\textit{uniform} sampling from $\mathcal{P}$ and pre-optimization with \textit{Sobol} sequences), and loss functions for the three best candidate systems while for each system the optimization is run ten times.}
    \label{tab:results}
\end{table}

Since the parameter optimization of Algorithm~\ref{alg:optimization} is repeatedly used for determining the fitness of a candidate graph, it needs to converge robustly. Thus, we investigate the effectiveness of the parameter initialization and the differences between loss functions in Table~\ref{tab:results}. We observe that our Sobol initialization leads to robustly learning parameters that achieve high satisfaction with the \ac{stl} specification and reach \ac{stl}-satisfying parameters set with fewer gradient steps. While the adaptive loss does lead to a lower mean robustness, it leads to higher additive robustness of subspecifications. The steps necessary for reaching the convergence criteria are similar when considering the standard deviation intervals. Additionally, we observe that with the loss $\mathcal{L}_\rho$, it happens that a gradient step leads to leaving the satisfaction parameter space again, especially for larger learning rates.

\section{Discussion} \label{sec:discussion}
While the numerical example above validates the approach,
further studies are needed for
broad applicability. In particular, the single initial state and noise-free input are a simplification of reality. The extension to state and input sets can be achieved by sampling trajectory batches from these sets and applying gradient descent to optimize the \ac{ode} parameters on the batch. Additionally, the assumption of a pre-defined set of species could be relaxed to uncertain species by iteratively applying our genetic algorithm while adding more species to the set of species referenced in the \ac{stl} specification until the specification is satisfied. The best network identified in our example does not have a path between the species on the left side of the implications to the species on the right side of the implications. If such behavior was intended, it could be forced by constraining the possible edge set $\mathcal{F}$.

We restricted the used \acp{ode} to Hill equations for activation and inhibition, and degradation terms. However, biological systems can include other reactions such as sequestration or logarithmic growth. Our graph representation could be extended by additional edge types for these reactions. This would result in an exponential increase of possible graphs, which could be challenging for the convergence of the genetic algorithm. Gradient-based methods for graph neural networks \citep{franceschi2020-gnninference, zheng2020-gnninference2} could potentially improve scalability, but they usually require many examples to learn. Another potential improvement is using advanced \ac{ode} solvers suited for stiff \acp{ode}, e.g.,  Kverno5 \citep{kvaerno2004-kverno}, instead of the Euler method. To achieve backpropagation through these solvers, techniques as proposed for Neural \acp{ode} could be used, e.g., the automatic differentiation framework or  the adjoint method \citep{chen2019-NeuralODEs}.

For systems with limited qualitative information, a large set of candidate networks can often comply with the given specifications due to underspecification (as in our example). When more information becomes available through experiments or expert feedback, the candidates could be refined so that eventually the real system model is inferred or the model that can be used for synthesizing a biological system in the lab is identified. 
Our approach is a flexible framework that can account for new information by refining the \ac{stl} specification or by biasing the population generation with the new knowledge or with previously found candidates.
When quantitative data becomes available, it can be directly incorporated by adding a loss term or using \ac{stl} inference methods, such as those presented by \cite{li2024-listlnn} or \cite{bombara2021-decisiontree} to convert the quantitative data into refined \ac{stl} statements.

\section{Conclusion} \label{sec:conclusion}
We proposed a learning approach for biomolecular models from \ac{stl} specifications of system behavior, 
which is suitable for inference and synthesis. 
We introduced a graph representation that is mapped to biologically-grounded \acp{ode} so that our genetic algorithm can simultaneously optimize the topology of the graph and the parameters of the \acp{ode}. We addressed the challenge of robust parameter initialization and choice of appropriate loss functions. On a numerical example, we observed that our approach identifies multiple system candidates that satisfy the specification. Thus, validating that our approach achieves learning of biomolecular models from \ac{stl} specifications, which describe qualitative observations. 

\acks{
This work was funded in part by the Air Force
Office of Scientific Research grant 
FA5590-23-1-0529, 
NSF GCR 2219101, and NIH R01 EB030946.
}

\vskip 0.2in
\bibliography{references}

\end{document}

%% file: figures/overview_fig.tex
\def\stepbox#1#2#3#4{
	\begin{scope}[shift={#1}, rotate=#2, scale=#4]
		\draw [draw=#3, rounded corners, very thick](-3,-5) rectangle (11,8.5);
	\end{scope}
}

\def\stepboxmid#1#2#3#4{
	\begin{scope}[shift={#1}, rotate=#2, scale=#4]
		\draw [draw=#3, rounded corners, very thick](-3,-5) rectangle (23,8.5);
	\end{scope}
}
\resizebox{\textwidth}{!}{%
	
\begin{tikzpicture}[node distance=4cm, auto]  
	\tikzset{
		mynode/.style={rectangle,rounded corners,draw=black,fill=TUMBlue!80,very thick, inner sep=1em, minimum size=3em, text centered, minimum width=10cm, text=white, minimum height=1.5cm},
		myenhancer/.style={->, thick, shorten <=2pt, shorten >=2pt,},
            myinhibitory/.style={-|, thick, shorten <=2pt, shorten >=2pt},
            mydashed/.style={draw=TUMGray, thin, dashed, dash pattern=on 3pt off 2pt}
		mylabel/.style={text width=7em, text centered},
            every state/.style={draw=black,text=black,minimum size=2cm,font=\Huge} 
	} 
	\begin{scope}[yshift=-7cm, xshift= -9cm]
       
		\draw [draw=TUMGray!40, rounded corners, very thick](-3,0.5) rectangle (11,8.5);
		\node[text=TUMGray!90] at (3.7, 9) (legend1){\Huge{ \makecell[c]{\textbf{Genes, cell or cytokine types}}}};
         \begin{scope}[yshift=1cm]
            \node[state] (A) at (0,5)   {1};
            \node[state] (B) at (4,4.5)   {2};
            \node[state] (C) at (2,1.5)  {3};
            \node[state] (D) at (6.5,2)   {4};
            \node[state] (E) at (7,5.5)   {5};
        \end{scope}
        \draw [draw=TUMGray!40, rounded corners, very thick](-3,-5) rectangle (11,-1);
		\node[text=TUMGray!90] at (3.7, -0.5) (legend2){\Huge{ \makecell[c]{\textbf{\ac{stl} specification}}}};
        \node[text=black] at (3.7, -3) (spec){\Huge{ \makecell[c]{$\varphi: x_1 \geq 0.4 \implies \event_{[5,10]} x_5 \leq 20 $}}};
        \begin{scope}[yshift=3cm]
		\draw [fill=TUMGray!30,draw=TUMGray!30, very thick, yshift=-0.8cm] (11.25,3) -- (14.7,3) -- (15.2,2.5) -- (14.7,2) -- (11.25,2)-- cycle;
		\draw [fill=TUMGray!30,draw=TUMGray!30, very thick, yshift=-0.8cm] (14.4,1.5) -- (15.5,2.5) -- (14.4,3.5)-- cycle;
        \end{scope}
        \begin{scope}[yshift=-4.5cm]
		\draw [fill=TUMGray!30,draw=TUMGray!30, very thick, yshift=-0.8cm] (11.25,3) -- (14.7,3) -- (15.2,2.5) -- (14.7,2) -- (11.25,2)-- cycle;
		\draw [fill=TUMGray!30,draw=TUMGray!30, very thick, yshift=-0.8cm] (14.4,1.5) -- (15.5,2.5) -- (14.4,3.5)-- cycle;
        \end{scope}
	\end{scope}
        \begin{scope}[yshift=-7cm, xshift= 9.8cm]
            \tikzset{
                every state/.style={draw=black,text=black,minimum size=1.5cm,font=\LARGE}
                }
		\stepboxmid{(0,0)}{0}{TUMGray!40}{1}
		\node[text=TUMGray!90] at (10.5, 9) (legend1){\Huge{ \makecell[c]{\textbf{Genetic algorithm}}}};
                \begin{scope}[scale=1,yshift=-0.5cm]
                \draw [draw=TUMGray!90, rounded corners, very thick](-0.0,1) rectangle (9,7.5);
                \draw [draw=TUMGray!90, rounded corners, very thick,fill=white](-0.5,0.5) rectangle (8.5,7);
                \draw [draw=TUMGray!90, rounded corners, very thick, fill=white](-1,0) rectangle (8,6.5);
                \node[state] (A) at (0,5)   {1};
                \node[state] (B) at (4,4.5)   {2};
                \node[state] (C) at (2,1.5)  {3};
                \node[state] (D) at (6.5,2)   {4};
                \node[state] (E) at (7,5.5)   {5};
                \path (C) edge [myenhancer, draw=TUMGray, thin, dashed] node [font=\Large] {\textcolor{gray}{?}} (A)
                          edge [myenhancer, draw=TUMGray, thin, dashed] node [font=\Large] {\textcolor{gray}{?}} (B)
                          edge [myenhancer, draw=TUMGray, thin, dashed] node [font=\Large] {\textcolor{gray}{?}} (D)
                      (D) edge [myinhibitory,  draw=TUMGray, thin, dashed] node [font=\Large] {\textcolor{gray}{?}} (B)
                      (E) edge [myinhibitory, draw=TUMGray, thin, dashed] node [font=\Large] {\textcolor{gray}{?}} (D);
                \draw[very thick, -latex] (4.5,7.7) -- (4.5,8.2) -- (16,8.2) -- (16,7.7);
                \draw[very thick, -latex] (4.5,-1.8) -- (4.5,-0.2);
                \node at (10.5, 8.5) (legend1) {{\fontsize{19pt}{18pt}\selectfont \makecell[c]{$\mathcal{G}_1, \mathcal{G}_2, \mathcal{G}_3$}}};
                \draw[very thick, -latex] (16,-0.2) -- (16,-1.8);
                \node at (14.5, -0.9) (legend1) {{\fontsize{19pt}{18pt}\selectfont \makecell[c]{$\rho^*_1, \rho^*_2,\rho^*_3$}}};
                \end{scope}
            \begin{scope}[scale=1,shift={(12,-0.5)}]
                \draw [draw=TUMGray!90, rounded corners, very thick](-0.0,1) rectangle (9,7.5);
                \draw [draw=TUMGray!90, rounded corners, very thick,fill=white](-0.5,0.5) rectangle (8.5,7);
                \draw [draw=TUMGray!90, rounded corners, very thick, fill=white](-1,0) rectangle (8,6.5);
                \node[text=black] at (3.7, 3.2) (legend1) {{\fontsize{19pt}{18pt}\selectfont \makecell[c]{$\begin{aligned}
                    & \theta_1^* = \argmax_{\theta \in \mathcal{P}} \rho(\mathbf x(\theta, \mathcal{E}_1), \varphi)  \\[7pt]
                    &\text{such that}: \\
                    & \mathbf{x}(\theta, \mathcal{E}) = \int_0^T \mathcal{D}_{\theta, \mathcal{E}}(x, a) \, dt \\
                    & \mathbf{x}(t=0) = \mathbf{x}_0
                \end{aligned}$}}};
            \end{scope}
            \draw [draw=TUMGray!90, rounded corners, very thick](-1,-4.5) rectangle (21,-2.5);
            \node at (10, -3.5) (legend1) {\huge{ \makecell[c]{Mutation \hspace{0.4cm}$\leftarrow$ \hspace{0.4cm} Selection \hspace{0.4cm} $\leftarrow$ \hspace{0.4cm} Fitness Evaluation}}};
            \draw [fill=TUMGray!30,draw=TUMGray!30, very thick, xshift=12cm, yshift=-0.8cm] (11.25,3) -- (14.7,3) -- (15.2,2.5) -- (14.7,2) -- (11.25,2)-- cycle;
		\draw [fill=TUMGray!30,draw=TUMGray!30, very thick, xshift=12cm, yshift=-0.8cm] (14.4,1.5) -- (15.5,2.5) -- (14.4,3.5)-- cycle;
	\end{scope}

        \begin{scope}[yshift=-7cm, xshift= 41cm]
		\stepbox{(0,0)}{0}{TUMGray!40}{1}
		\node[text=TUMGray!90] at (3.7, 9) (legend1){\Huge{ \makecell[c]{\textbf{Optimized biomolecular network}}}};
        \begin{scope}[xshift=-1cm, yshift=0.5cm]
            \node[state]         (A) at (0,4)   {1};
            \node[state]         (B) at (4,3)   {2};
            \node[state]         (C) at (2,-1)   {3};
            \node[state]         (D) at (7,0)  {4};
            \node[state]         (E) at (6,6)   {5};
            
            \path (A) edge [myinhibitory, draw=TUMRed,thick] node [font=\Large] {\textcolor{gray}{$K_{15}, v_{5}$}} (E)
                  (C) edge [myenhancer, draw=TUMBlue,thick] node [font=\Large] {\textcolor{gray}{$K_{31}, v_{1}$}} (A)
                      edge [myenhancer, draw=TUMBlue,thick] node [font=\Large] {\textcolor{gray}{$K_{32}, v_{2}$}} (B)
                      edge [myenhancer, draw=TUMBlue,thick] node [font=\Large] {\textcolor{gray}{$K_{34}, v_{4}$}} (D)
                  (D) edge [myinhibitory, >=|, draw=TUMRed,thick] node [font=\Large] {\textcolor{gray}{$K_{42}, v_{2}$}} (B);
        \end{scope}
        \draw[myinhibitory, draw=TUMRed,thick] (7,7) -- (8,7);
        \node at (9.2,7) {\Large{inhibition}};
        \draw[myenhancer, draw=TUMBlue,thick] (7,7.5) -- (8,7.5);
        \node at (9.2,7.5) {\Large{activation}};
        \node at (4,-3) {\Huge{$\rho(\mathbf x(\theta^*, \mathcal{E}^*), \varphi) > 0 \quad \quad$ \textcolor{TUMGreen}{$\checkmark$}}};
	\end{scope}
    
\end{tikzpicture} 
}

%% file: figures/result_graph.tex
\resizebox{0.35\textheight}{!}{%
\begin{tikzpicture}[node distance=4cm, auto]  
	\tikzset{
		mynode/.style={rectangle,rounded corners,draw=black,fill=TUMBlue!80,very thick, inner sep=1em, minimum size=3em, text centered, minimum width=3cm, text=white, minimum height=1.5cm},
		myenhancer/.style={->, thick, shorten <=2pt, shorten >=2pt,},
            myinhibitory/.style={-|, thick, shorten <=2pt, shorten >=2pt},
            mydashed/.style={draw=TUMGray, thin, dashed, dash pattern=on 3pt off 2pt}
		mylabel/.style={text width=7em, text centered},
            every state/.style={draw=black,text=black,minimum size=2cm,font=\Large} 
	} 
            \node[state, fill=PLOTCyan!30]         (E4) at (0,5)   {1};
            \node[state, fill=PLOTDarkBlue!30]         (F5) at (0,0)   {2};
            \node[state, fill=PLOTDarkOrange!30]         (B1) at (5,5) {3};
            \node[state, fill=PLOTRed!30]         (D3) at (5,0)  {4};
            \node[state, fill=PLOTGray!30]         (A0) at (10,5)   {5};
            \node[state, fill=TUMgray!30]         (C2) at (10,0)   {6};

            \path 
            (E4) edge [myenhancer, draw=TUMBlue,thick, bend right] node [font=\Large, near start, left] {\textcolor{black}{$K_{12} = 0.36$}} (F5)
            (B1) edge [myinhibitory, draw=TUMRed,thick] node [font=\Large, left, near start] {\textcolor{black}{$K_{34} = 0.03$}} (D3)
            (D3) edge [myinhibitory, draw=TUMRed, thick] node [font=\Large] {\textcolor{black}{$K_{41} = 0.93$}} (E4)
            (D3) edge [myenhancer, draw=TUMBlue,thick] node [font=\Large] {\textcolor{black}{$K_{42} = 0.18$}} (F5)
            
            (A0) edge [myenhancer, draw=TUMBlue,thick] node [font=\Large] {\textcolor{black}{$K_{53} = 1$}} (B1)
            (A0) edge [myenhancer, draw=TUMBlue,thick] node [font=\Large, near start] {\textcolor{black}{$K_{52} = 0.43$}} (F5)
            (A0) edge [myenhancer, draw=TUMBlue,thick] node [font=\Large] {\textcolor{black}{$K_{56} = 0.69$}} (C2);
    
\end{tikzpicture} 
}